\documentclass[twocolumn,showkeys,showpacs,superscriptaddress,notitlepage]{revtex4-1}

\usepackage{graphicx}
\usepackage{amsmath,amsfonts,amssymb}
\usepackage{morefloats}
\usepackage{color}

\begin{document}

\title[ ]{Social imitation vs strategic choice, or consensus vs cooperation\\ in the networked Prisoner's Dilemma}

\author{Daniele Vilone}\email{daniele.vilone@gmail.com}
\affiliation{LABSS (Laboratory of Agent Based Social Simulation),
Institute of Cognitive Science and Technology,
National Research Council (CNR), 
Via Palestro 32, 00185 Rome, Italy}

\author{Jos\'e J.\ Ramasco}\email{jramasco@ifisc.uib-csic.es}
\affiliation{Instituto de F\'{\i}sica Interdisciplinar y Sistemas Complejos IFISC (CSIC-UIB), 07122 Palma de Mallorca, Spain}

\author{Angel S\'anchez}\email{anxo@math.uc3m.es}
\affiliation{Grupo Interdisciplinar de Sistemas Complejos (GISC), Departamento de Matem\'{a}ticas, Universidad Carlos III de Madrid, 28911 Legan\'{e}s, Madrid, Spain}
\affiliation{Instituto de Biocomputaci\'{o}n y F\'{i}sica de Sistemas Complejos (BIFI), Universidad de Zaragoza, 50018 Zaragoza, Spain}

\author{Maxi San Miguel}\email{maxi@ifisc.uib-csic.es}
\affiliation{Instituto de F\'{\i}sica Interdisciplinar y Sistemas Complejos IFISC (CSIC-UIB), 07122 Palma de Mallorca, Spain}

\begin{abstract}
The interplay of social and strategic motivations in human interactions is a largely unexplored question in collective social phenomena. Whether individuals' decisions are taken in a pure strategic basis or due to social pressure without a rational background crucially influences the model outcome. Here we study a networked Prisoner's Dilemma in which decisions are made either based on the replication of the most successful neighbor's strategy (unconditional imitation) or by pure social imitation following an update rule inspired by the voter model. The main effects of the voter dynamics are an enhancement of the final consensus, i.e., asymptotic states are generally uniform, and a promotion of cooperation in certain regions of the parameter space as compared to the outcome of purely strategic updates. Thus, voter dynamics acts as an interface noise and has a similar effect to a pure random noise; furthermore, its influence is mostly independent of the network heterogeneity. When strategic decisions are made following other update rules such as the replicator or Moran processes, the dynamic mixed state found under unconditional imitation for some parameters disappears, but an increase of cooperation in certain parameter regions is still observed. Comparing our results with recent experiments on the Prisoner's Dilemma, we conclude that such a mixed dynamics may explain moody conditional cooperation among the agents.
\end{abstract}



\maketitle

\section{Introduction}

Collective social phenomena arise from interactions of individuals, considered the elementary units of social structures \cite{cas09}. Related phenomena include, e.g., strategic interactions, opinion/cultural dynamics, epidemic and rumor spreading, and so on. In many of these problems, dynamics is not uniquely defined as in a Hamiltonian system, and it is often the case that social contexts can evolve in time in many different manners \cite{klu00}. This is particularly relevant in strategic interactions, in which people take into account their expectations about what their partners might choose to do \cite{sig10,gin09}. 

A specific and socially important problem where the dynamics leads to very different outcomes is the emergence of cooperation \cite{axe81,may95,ham03,kap06}. Indeed, the spreading and prosperity of cooperative behaviors, often observed in human societies despite their disadvantage in terms of fitness of single individuals, has been for years a central topic in Evolutionary Dynamics, Economics, Sociology and Psychology. Several models, mechanisms and ideas have been suggested in order to explain the emergence and the stability
of cooperation \cite{feh03,now06,fle09}, particularly in games where cooperation is costly such as Prisoner's Dilemma (PD). One such mechanism is the so-called network reciprocity, which was first proposed by 
Nowak and May~\cite{now92}. They showed that if the PD is played on a lattice frozen configurations can appear, where cooperators have survived and even overwhelmed defectors. Subsequent research established that the effect of the topology is not universal \cite{roca09b} (see also \cite{rcs09} for a recent review), and the enhancement of the cooperative strategies depends on the network properties, the dynamical update rule, the exact entries
of the payoff matrix, the size of the system and the time structure~\cite{hub93,now04,egu05,san06,roc09,jia09,vil11}. The theoretical discussion was finally settled by several experiments \cite{kir07,tra10,gru10,gra12b,gru14}, which established that networks did not exhibit a cooperation level significantly higher than a well mixed population. 
Interestingly, those experimental works allow to conclude that most of the update rules that have been used in analytical or numerical studies of the evolutionary PD are not used by actual people when playing the game, and therefore other dynamics must be considered. 

In this paper, we propose an evolutionary update rule that combines strategic thinking with imitation of a more social character, and study how it affects the outcome of the networked PD. Indeed, 
it has been noticed that in the real world a key point is that people do not reach a decision
only on the basis of a strategic reasoning, but also considering the social pressure of their environment, with the possibility
of making mistakes~\cite{gra78,yu011}.
The first and most common way for an individual to follow social pressure is imitating a neighbor's act or opinion without any strategic considerations behind. To represent this, in a previous work~\cite{vrss12} we introduced a model where agents can evolve by a mixed dynamics of pure and strategic imitation in a coordination game, finding that such mixing makes possible that the system orders in one of the two absorbing states in situations in which neither the pure coordination game nor the voter model reach consensus. Our aim here is to broaden the topic studying what happens when pure imitation acts together with strategic dynamics on the PD, verifying if also in this case the non-strategic behavior leads to a more convenient final configuration for the whole society. In addition, we will compare our results to the available experimental evidence, drawing conclusions as to the relevance of the proposed dynamics to the real world. 

The paper is organized as follows: in Sec.\ II, we present in detail our model and the rationale behind its definition. Section III deals then with the numerical results obtained in two relevant systems, namely random (Erd\"os-R\'enyi, ER) and scale free (SF) networks. Subsequently, in Sec.\ IV, it is analyzed how the system outcome changes when other update dynamics are used in the strategic decisions. Section V discusses our work on the light of the experiments and finally Sec.\ VI summarizes our main findings. 

\section{Model description}  

Our model consists of a system of $N$ agents on a network, which for the purpose of this work will be either Erd\"os-R\'enyi or random SF (generated with the configurational model) as paradigmatic examples to explore the effects of heterogeneity in the number of connections. Additionally, experiments on the behavior of human subjects have been carried out on the same type of inhomogeneous networks \cite{gra12b}, which facilitates the comparison between our results and the experimental ones. Each agent interacts directly only with her nearest neighbors in the network, and can choose among two possible actions, represented 
 by a two-valued integer variable (``action'') $s=\pm1$. We identify the positive value with cooperation ({\bf C}) and the negative one
with defection ({\bf D}). Individuals interact in a PD, i.e., they play with their neighbors and then collect a payoff
according to the action adopted by themselves and their opponents. Payoffs are collected according to the following payoff matrix: 
\begin{equation}
\begin{tabular}{|c|c|c|}
\hline
\mbox{ } & {\bf C} & {\bf D} 
 \\ \hline
{\bf C} & $1$ & $0$
 \\ \hline
{\bf D} & $T$ & $\varepsilon$ \\
\hline
\end{tabular}
\label{3b} 
\end{equation}
where the punishment parameter $\varepsilon$ must fall in the interval $[0,1)$ and $T>1$ to ensure that the game is a PD. Without loss of generality, we set $T=1.4$ and vary $\varepsilon$ in our simulations. Note that this is not the usual way to work with the PD in most papers, as typically $\varepsilon$ is set to $0$ and the payoff to C vs D is in a range of zero or negative values. The reason for this particular choice is to facilitate the interpretation of our results on the light of earlier work of ours on symmetric coordination games \cite{vrss12}, given by the same payoff matrix as in Eq.\ (\ref{3b}) with $T=0$ and $\epsilon = 1$. In the game considered here, it is  well known that the rational choice is always to defect, not only to prevent betrayals by other players but also because it always gives higher payoff no matter what the other player does (in economic terms, it is a Nash equilibrium \cite{gin09}). However, it is also clear that the global gain would be higher if both agents cooperate, and just in here lays the dilemma. 

Having defined our strategic context, we now turn to the dynamics. After every interaction, 
an individual updates her strategy simply imitating a neighbor at random, that is, following the voter model (VM)
dynamics with probability $q$. Otherwise, with probability $1-q$, an update rule based on the actual fitness attained by herself and her neighbors is implemented. For the main part of the paper, we will use unconditional imitation (UI) as the strategic update rule.  
With UI, at the end of each round of the game, every player
imitates the strategy of the neighbor that has obtained the best payoff provided it is
larger than her own. This update rule was introduced in \cite{now92} as a simple way to implement the interest of a (rational) agent to maximize her own fitness and to reach that end by 
imitating the most successful individuals. It is important to note that there are other strategic imitation rules that could be considered, some of which are discussed in Sec.\ IV below as alternatives to our main model. Experimental results, however, show that UI might resemble the actions of the players in real life, provided they do not stick to the update rule all the time \cite{gru10,gru14}. This is a further motivation to consider a mixed dynamics as we are proposing here. On the other hand, regarding the VM, in spite of its simplicity it has been shown to capture some features of the way people behave in, e.g., electoral processes \cite{fsrse13} and therefore it is a very suitable manner to introduce mechanisms of social imitation in an otherwise strategic problem. 


\begin{figure}
\begin{center}
\includegraphics[width=8.6cm]{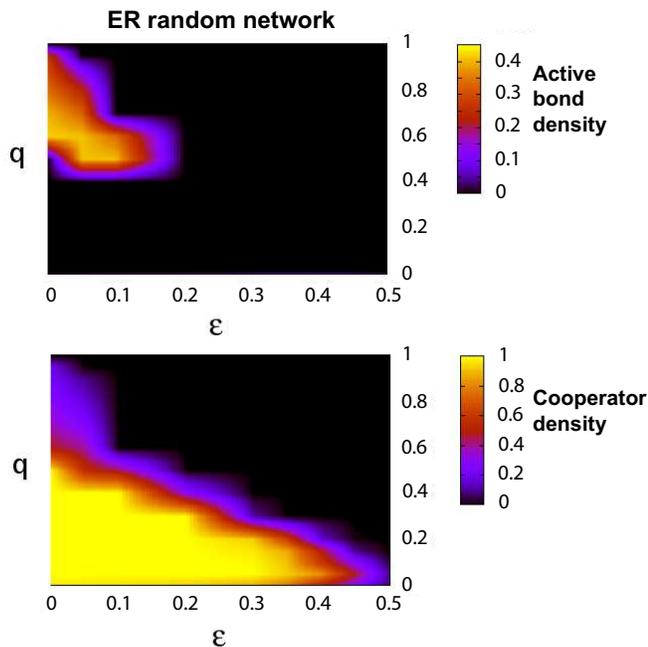}
\caption{Final active bonds density (top) and cooperator density (bottom) as a function of the dynamics mixing parameter $q$ and the punishment $\varepsilon$ for a system on an ER random network of size $N=3000$ and average degree $\langle k\rangle= 8.48$. Caveat: the interval of parameter space investigated is
$(\varepsilon,q)\in [0,0.5]\times(0,1)$, notice that $q=0$ and $q=1$ are excluded.}
\label{SFPDG_3Dnc_bis}
\end{center}
\end{figure}

For reference, it is useful to recall what are the effects of either of our dynamics when acting alone.  
In high dimensional lattices ($d \ge 3$) and random networks, the VM dynamics drives the system to a disordered active state, whose proportion of opinions (in our case, actions) is given by the initial conditions~\cite{cas09,ses05}. This state is dynamic in the ensemble average, the opinion of the agents keep changing along time. In finite systems, the fluctuations will eventually drive the system to consensus, which is an absorbing state. The characteristic time to reach consensus grows with system size $N$, diverging in the limit $N \to \infty$. As for the strategic dynamics, while in mean field PD ends up in a consensus (frozen) state with all defectors regardless of the update rule, in complex topologies it is possible that some cooperators survive, even though this effect is not universal and the final state depends sensitively on an entire set of
different parameters. A full summary of the different outcomes can be found in the review \cite{rcs09}. For our basic rule, UI, the outcome of the evolution on random networks is in general also full defection, but in a small parameter region around $\varepsilon=0$ cooperation prevails, the more so the lower the degree of the network. Such promotion of cooperation in a bounded range of parameters takes place as well for scale free networks, although in this case the cooperative region is smaller than for random networks. Note, however, that (whether ordered or disordered) the final outcome of the system on the random networks with strategic rules is in general a frozen state, where no player changes her decision anymore.  Finally, it is also interesting to mention what happens when the interaction is given by a coordination game \cite{vrss12}. In this case, as explained, the final outcome of either rule (VM or UI) alone is a disordered state active or frozen but the mix of both UI and VM completely changes the scenario because the system tends to consensus either rapidly when the VM  is dominant or slowly when it acts as a noise over the game-like evolution.

\section{Numerical results}

We begin the report of our numerical results by checking whether the mixing of the two dynamics influences how the system reaches the final state, and the nature of the final state itself. Unless otherwise stated, we will be presenting results averaged over $1000$ system realizations. Figure \ref{SFPDG_3Dnc_bis} summarizes our results for an ER random network with $\langle k\rangle= 8.48$. In the top plot, the final value of the active link density is shown as a function of the punishment and of the mixing parameter. The active link density is the fraction of links connecting nodes with opposite strategies. In the case of VM dynamics, these links are susceptible of further updates. As can be seen in the bottom panel, for a majority of parameter choices the system becomes frozen in a consensus state (either all C or D). Nevertheless, there exists a parameter region in which the fate of the system is not frozen but a dynamic state. As mentioned above, this is not the case when the game is of coordination, and therefore it is a new phenomenon arising for the PD. This region corresponds to low values of $\varepsilon$, meaning that the incentive to defect is relatively small, and high values of $q$, i.e., large probability of imitating a neighbor at random (VM). Therefore, the dynamics is such that as the transition from cooperator to defector is relatively slow due to the strategic update there remains a pool of cooperators than can be socially copied and thus an overall intermediate level of cooperation is maintained. In the regions where full consensus is reached,  the cooperator density in the plot is actually the probability that the final configuration of a realization is a full C state. An interesting feature demonstrated by these plots is that cooperation is enhanced with respect to the classical mean-field case and, although the comparison is not straightforward due to the different parameterization, it is also enhanced when compared with pure UI dynamics on a random network. 

\begin{figure}
\begin{center}
\includegraphics[width=8.6cm]{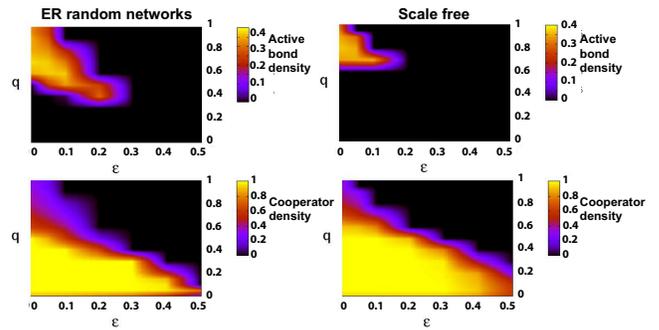}
\caption{Final active bonds density (top) and cooperator density (bottom) as a function of the mixing parameter $q$ and $\varepsilon$ for a system on networks of size $N=3000$. The network is an ER on the left and scale free on the right with average degree $\langle k\rangle=5.14$. Caveat: the interval of parameter space investigated is $(\varepsilon,q)\in [0,0.5]\times(0,1)$, notice that $q=0$ and $q=1$ are excluded.}
\label{PDG_3Dna}
\end{center}
\end{figure}

As these results are obtained for a specific type of networks, it is important to check their generality if the average degree or the full network topology is changed. For comparison,  Figure \ref{PDG_3Dna} presents the final value of the active link density and the
cooperator density for an ER random network and for a scale free network of average degree  $\langle k\rangle=5.14$ . It can be seen that decreasing the number of neighbors leads to an expansion of the region where cooperation is observed, in agreement with what takes place with UI dynamics only; as the number of neighbors decreases it becomes more likely to imitate a cooperator and the network may be driven towards full cooperation. On the contrary, the case of the scale free network behaves differently: adding more heterogeneity to the network leads to a slightly larger region of cooperative behavior and, at the same time, to a smaller region where the dynamics leads to an active state. Notwithstanding, the general conclusion one can draw from these plots is that the network properties, at least in the class of uncorrelated networks we are looking at, do not affect very much the outcome of the mixed dynamics.

\begin{figure}
\begin{center}
\includegraphics[width=8.6cm]{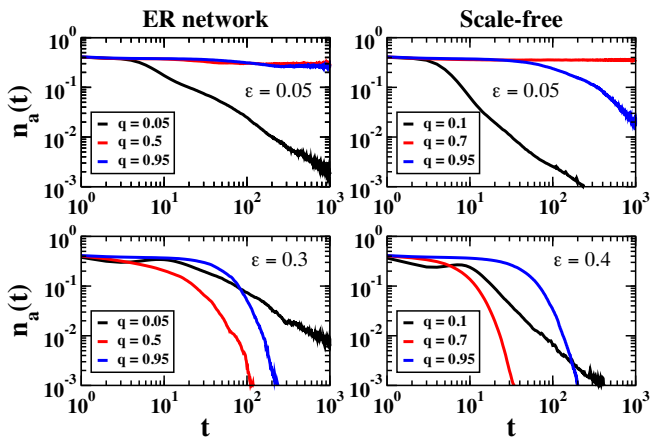} 
\caption{Time evolution of the average active links for a system on networks of size $N=3000$; a random network  on the left and 
a scale free network on the right, both with $\langle k\rangle=5.14$. Top figures are with low value of $\varepsilon$ (0.5 in both
cases); bottom ones are with higher punishment parameter (0.3 and 0.4, respectively).}
\label{erui-eps0.30}
\end{center}
\end{figure}

\begin{figure}[b]
\begin{center}
\includegraphics[width=8.6cm]{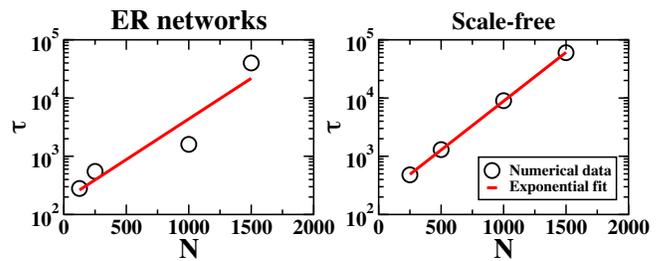} 
\caption{Dependence of the time needed to reach the final state  $\tau$ as a function of the system size $N$. Operatively, $\tau$ has
been measured as the time when the active link density dropped below $0.07$. The networks have a random (left) or scale free (right) topology with average degree $\langle k\rangle=5.14$. The  punishment has been set at $\varepsilon=0.05$ and $q=0.5$ and 0.7, respectively. In the exponential fits $\tau\sim\exp(\gamma N)$, we have estimated $\gamma\simeq0.0032\pm0.0008$ for the random networks and $\gamma\simeq0.0038\pm0.0001$ for the scale free networks.}
\label{tau_vs_L}
\end{center}
\end{figure}

Let us focus now on the details of the time evolution of the system. 
Figure \ref{erui-eps0.30} shows examples of  the behavior of the average active links density in a system
evolving on a random or on a scale free network for different values of the punishment
$\varepsilon$. 
For larger values of the punishment,
we find that the system always reaches a consensus state, following an exponential decay for high values of $q$, and a power-law-like decay for small values of the mixing parameter. This behavior is qualitatively the same in both types of networks, and it is in good agreement with the results
of the model studied in~\cite{vrss12}, where players interacted through a coordination game. 
When the punishment is small,  for small $q$ the system still reaches a consensus state. However  for higher $q$ it ends up in a dynamic state. It is interesting to stress that for the $q$ values chosen in the case of the scale free network convergence to consensus or an active mixed state is in fact a reentrant phenomenon, as was expected from the results in Figure~\ref{PDG_3Dna}. The fact that the system reaches a real active final state
is proven in Figure~\ref{tau_vs_L}, where the time needed to reach consensus, $\tau$, is shown to diverge exponentially with the system size. In the thermodynamic limit, the system never orders.

\begin{figure}
\begin{center}
\includegraphics[width=8.6cm]{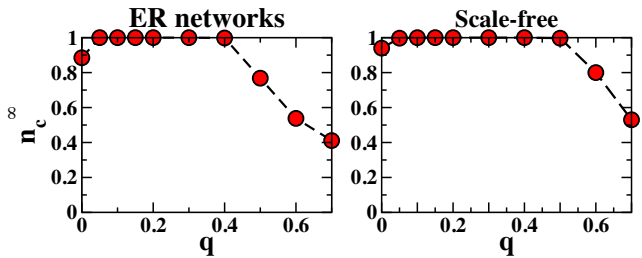} 
\caption{Final cooperator density for a system on a random (left) or scale free (right) network of size $N=3000$ and average degree $\langle k\rangle=5.14$. The punishment parameter is  $\varepsilon=0.05$. $q = 0$ means a purely strategy update of the PD, note that there is a slight increase of $n_c^\infty$ as soon as $q$ becomes larger than zero. }
\label{PDG_nc1}
\end{center}
\end{figure}

To gain further insight on the properties of the final state, we check whether cooperation is promoted
or not by the influence of the VM dynamics. The final cooperator density is shown in Figure~\ref{PDG_nc1} as a function of $q$ for $\varepsilon=0.05$. The aim is to provide a clearer picture than the general phase diagrams $(\varepsilon,q)$ discussed above. Even though in absence of any other model contribution ($q = 0$) the cooperation
is already promoted (as compared to mean field) due to topological effects~\cite{rcs09}, we can notice how initially the mixing of dynamics actually enhances cooperation compared with the case of pure PD game. Subsequently, upon increasing $q$, the final cooperator
density begins its decay towards zero (not shown), apart from the singular limit at $q=1$ where, as already stated, only the voter
dynamics is acting and the system remains in an active disordered state (more precisely, the limit
of $n_c^{\infty}$ for $q\rightarrow1^-$ is zero, but an infinite time is needed to reach it). Once again, the behavior is similar on the two types of networks studied. This is reminiscent of the promotion of cooperation due to random noise (or mutation) analyzed in \cite{rcs09b}. However, the two models are different because in the case of \cite{rcs09b} noise acted upon any network node, through a random change in the current action of the player. On the contrary, what we have here is interfacial noise, because it only acts through imitation of neighbors who are playing a different action. Therefore, the noise introduced by the VM does not act on players who are completely surrounded by others with their same action. Notwithstanding, we do observe that an appropriate amount of VM leads to an increase of the cooperation.

\section{Other update rules}

\begin{figure}
\begin{center}
\includegraphics[width=8.6cm]{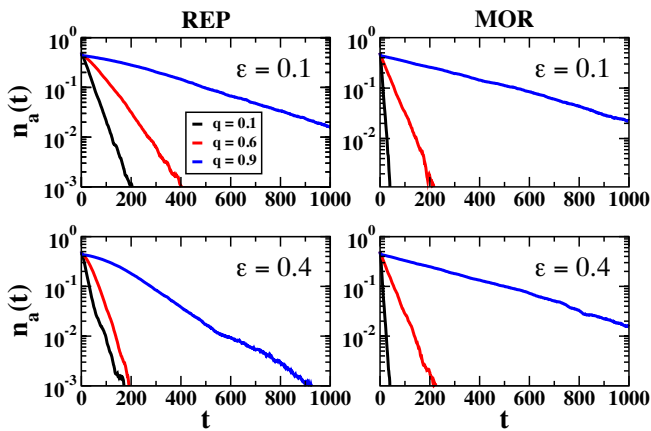} 
\caption{Time evolution of average active links for a system on a random network of size $N=3000$ and average
degree $\langle k\rangle=8.48$ with punishment $\varepsilon=0.1$ (top) and $\varepsilon=0.4$ (bottom). Left: REP update rule. Right: MOR update rule.}
\label{errep-eps0.40}
\end{center}
\end{figure}

To complete our study and enlarge the understanding of the combined effects of social imitation and strategic dynamics, we will now consider two additional dynamics, namely, replicator
(REP) and Moran (MOR) rules. With REP agents choose a neighbor at random: if the payoff of
the chosen neighbor is lower than the agent's own, nothing changes, but if it is larger,
the agent will adopt the neighbor's strategy with a probability proportional to the
difference between the two payoffs. On the other hand, with MOR, agents choose one of their neighbors with probability proportional to their payoff, without considering whether it is larger than hers or not. We chose these two rules because they complement our study of UI: indeed, UI is a local deterministic
rule (an agent watches all her neighborhood and the evolution is completely predetermined by the rule), REP is pairwise and stochastic (an agent decides how to evolve watching only one neighbor per time, and the result of the evolution is not univocally determined), MOR is local and stochastic. Moreover, MOR allows the individuals to make mistakes (there is a non-zero probability to imitate a neighbor with worse fitness). In fact, MOR can be understood as a weighted social imitation, where the weight is the success of the individual observed. We stress, however, that these two rules have not been observed in experiments and therefore their interest here arises from the viewpoint of the understanding of the mechanisms of the mixed dynamics. 

\begin{figure}
\begin{center}
\includegraphics[width=8.6cm]{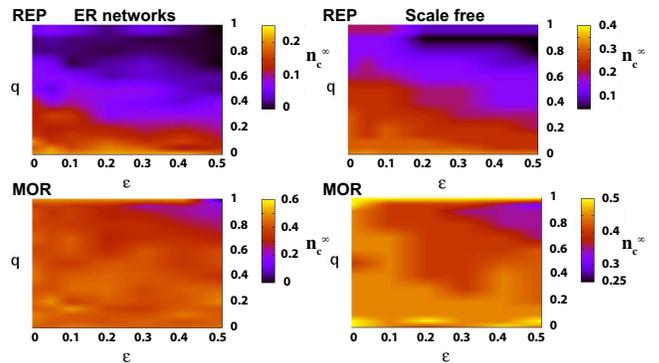}
\caption{Final cooperator density as a function of $q$ and $\varepsilon$ for a system on a random (left) or scale free (right) network of size $N=3000$ and average degree $\langle k\rangle=8.48$ and $\langle k\rangle=5.14$, respectively. Top: REP update rule. Bottom: MOR rule. Caveat: the interval of parameter space investigated is $(\varepsilon,q)\in [0,0.5]\times(0,1)$, notice that $q=0$ and $q=1$ are excluded.}
\label{PDG_3DncREP}
\end{center}
\end{figure}

The first important difference when the update rule changes is the disappearance of the final active state for any value of the parameters. In all the $(\varepsilon,q)$ phase space, the behavior is always as in the examples depicted in Figure~\ref{errep-eps0.40}: the active link density vanishes and a final consensus is reached, with all the network choosing the same action (either C or D). Moreover, the decay is in all cases exponential, pointing to the power-law behavior observed with UI as a specificity of that algorithm. In practice, what we find is that the stochasticity of the strategic evolution algorithm speeds up the dynamics and helps the system reaches always final consensus, whereas the pairwise or local character of the rule is less relevant.

\begin{figure}
\begin{center}
\includegraphics[width=8.6cm]{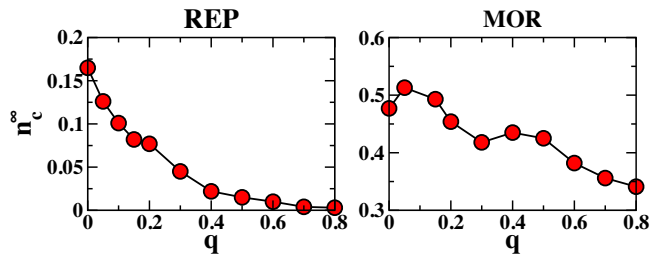}
\caption{Final cooperator density for a system on a random network of size $N=3000$ and average
degree $\langle k\rangle=8.48$. The punishment parameter is $\varepsilon=0.10$. Left: REP update rule. Right: MOR rule. Note that since the final configuration is static $n_c^\infty$ corresponds to the fraction of realizations ending in full cooperation.}
\label{PDG_MOR}
\end{center}
\end{figure}

As with the UI strategic dynamics, the type of network considered (within the broad class of uncorrelated networks of high dimensionality) is not relevant, in particular with regard to the final configuration. MOR rule enhances cooperation with respect to REP, as shown in
Figure ~\ref{PDG_3DncREP}, and this increase is always somewhat larger in scale free networks. It can be noticed that while for REP the final cooperator density drops
rapidly with increasing $q$, for agents evolving with the Moran rule the final cooperator density falls more slowly and in a longer range of $q$ values (Fig.~\ref{PDG_MOR}). In practice, while with MOR rule
inserting non-strategic imitation slightly enhances (or at least does not hinder) cooperation just as with UI, when we have REP the voter dynamics appears to hamper
the spreading of the cooperative strategy. As indicated above, MOR can be seen as a generalized voter, and the result is that the dynamics is similar in the sense that it leads with similar probability to any of the two possible outcomes. The final system aftermath is slightly biased towards the equilibrium, but, interestingly, always in a frozen configuration, which does not occur with VM alone. Finally, it is clear that in both cases there is only a weak dependence on $\varepsilon$.

\section{Discussion}

This far, we have studied the combination of strategic and social imitation dynamics as driver for evolution in a networked PD. We have thus learned that when strategic updates obey the UI rule, the mixing of  dynamics helps the system to reach consensus in an all C or D configuration. Note that for certain values of the parameters the final state is full cooperation, which is not an equilibrium of the underlying game. Furthermore, there is a small parameter region (high $q$ and low $\varepsilon$) where the final fate of the system is a dynamic state with a fraction of cooperators and defectors.

\begin{figure}
\begin{center}
\includegraphics[width=8.6cm]{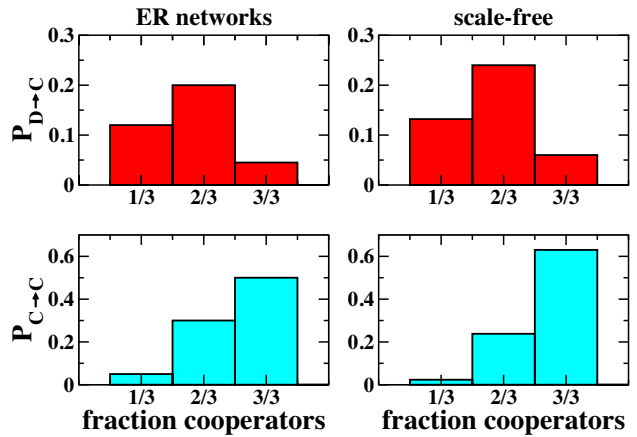}
\caption{
Evidence for moody conditional cooperation on systems given by 
a random network (left) and a scale free network (right) of size $N=3000$ and average degree $\langle k\rangle=5.14$.  Top: frequency of cooperation after defection as a function of the fraction of cooperating neighbors (see text) in the previous round. Bottom: frequency of cooperation after cooperation as a function of the fraction of cooperating neighbors in the previous round. Note that the histograms are similar regardless of the network topology. The payoff matrix is the same as in \cite{gru10} for comparison, with $\varepsilon=0$, $T=1.4$, and the payoff to a cooperator facing a defector vanishing as well; UI update rule.}
\label{MCC_SF_ER}
\end{center}
\end{figure}

In view of these results, a first question arises as whether some analytical explanation of those observations is possible. Unfortunately, the approach used for this purpose in~\cite{vrss12} does not work here due to the strongly different nature of the two games: VM and PD do not have the same equilibria as VM and the coordination game. Indeed, the outcome of the evolution for the coordination game is always a Nash equilibrium, and when the PD converges to full cooperation, it is not. There is, however, a possibility to address the issue in a limited parameter regime by considering a random walk in a weighted directed network.
It is known that the convergence time of the voter model has the same distribution
as the convergence time of the coalescing random walk process to a single particle~\cite{ald83,lig05,yil10},
which in its turn is related to the mean first passage time (MFPT) of a single walker~\cite{noh04}.
Then, for $q=1$ (i.e., pure VM dynamics) our problem is exactly that of the random walk on complex networks.
On the other hand, for $q<1$, because of the evolutionary rules, the agents with higher fitness can be seen as nodes where the probability of a walker to fall into increases and, at the same time, it becomes more difficult to escape from them. Now, for an undirected network, the MFPT from whatever starting point to an arrival node $j$, $\tau^u_j$, is~\cite{noh04,new04}
\begin{equation}
\label{Tu}
\tau^u_j=\frac{D}{\sum_l A_{lj}} ,
\end{equation}
where $A_{lj}$ is the adjacency matrix, which includes the weight of each link, and $D$ is an appropriate constant. Considering a situation with a mixing parameter $q$ close to one, let us say, $q=1-\eta$ with $\eta$ small enough, the MFPT can be approximated by  
\begin{equation}
\label{Td}
\tau_j = \tau_j^u+\alpha_j(\eta) ,
\end{equation}
where $\alpha_j$ should be a suitable perturbation depending on $\eta$ and on the node $j$. Then, if $j$ is occupied by an individual with the best fitness of all her neighborhood and having in mind that the strategic evolution is UI, no walker will ever reach node $j$, {\it i.e.} $\alpha_j\rightarrow\infty$, then also $\tau_j$ diverges. For the other rules we have checked, with MOR rule
$\alpha_j$ always remains finite, whereas with REP rule the system behaves in principle like UI, but since it hinders
considerably cooperation after relatively few time steps, it is easy for the finite system to undergo a
fluctuation (due to the voter dynamics) which let it land at the all-defectors consensus.
Of course, such argument is only qualitative and does not
fully explain what goes on to the dynamics for every value of $q$ close to $1$, but at least provides some 
justification of our observations under an UI update rule. 

\begin{figure}
\begin{center}
\includegraphics[width=8.6cm]{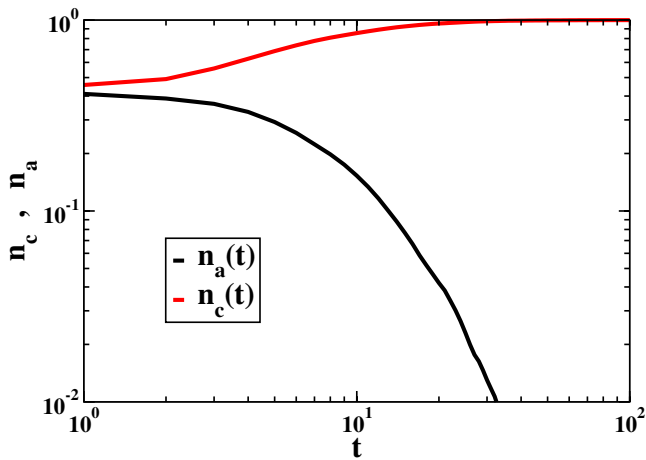}
\caption{
Time evolution of the fractions of cooperators and active links on a scale free network of size $N=3000$ and average degree $\langle k\rangle=5.14$. The payoff matrix is the same as in \cite{gru10} for comparison, with $\varepsilon=0$, $T=1.4$, and the payoff to a cooperator facing a defector vanishing as well; UI update rule.} 
\label{MCC_convergence}
\end{center}
\end{figure}

Beyond finding analytical insights on our results, the other important question that requires discussion is the relation of this dynamics with experiments. The experiments on square lattices~\cite{gru10,gru14} or on heterogeneous networks \cite{gra12b} show that subjects behave in a way that has been called moody conditional cooperation (MCC). This means that the action they take depends on the number of cooperating neighbors they had in the previous round: the more neighbors cooperated, the more likely it is that the player cooperates. However, the choice depends also on the player's own previous action: thus, it has been shown that cooperation following cooperation is much more likely than following defection. We have therefore checked whether the dynamics of our simulations is compatible with this observation. To this end, we looked at the probability of cooperation of an agent as a function of her last action and depending on the fraction of cooperating neighbors the last time she was selected to play a round. Also, in order to get better statistics and compare the results, we consider three ranges of cooperation contexts: when less than one third of the neighbors cooperated, when more than two third cooperated, and the intermediate case. We then estimate the probabilities of cooperation by looking at the frequencies with which the corresponding actions are chosen. 

Our results are shown in Figure \ref{MCC_SF_ER}, where it can be seen that we indeed observe a behavior of the agents reminiscent of the experimental observations. Indeed, we see that after defecting the probability to cooperate in the next round is low and not very dependent on the context, while after cooperating there is an almost linear dependence of the probability to cooperate again on the number of neighboring cooperators, as reported in the experiments \cite{gru10,gra12b,gru14}. Moreover, again in accordance with experiments, there is no dependence on the topology. The simulation has been done with a value of $q=0.3$, which we have chosen seeing that in the experimental data, if there is any imitation of the best neighbor, it happens at most in around $70\%$ of the cases. However, it is important to stress that these measurements are carried out during the transient phase: as we already stated, for most of the parameter space the final state of the system is either full cooperation or full defection,  and in particular for payoff values similar to those used in the experiments, the system converges to full cooperation (see Fig. \ref{MCC_convergence}). This is not in agreement with the experimental observations, where, if there is indeed convergence to a homogeneous state, it is very slow and much more likely towards full defection. In previous works \cite{vrss12,gar12}, we have observed a much slower dynamics for the cooperation fraction but associated to other type of games or more involved opinion models. On the other hand, we have checked that the same behavior arises for a very wide range of values of $q$ and $\varepsilon$, and hence in this sense the relation between moody conditional cooperation and our mixed dynamics is quite robust. In any event, we stress that such mixing of dynamics is not a necessary condition to get MCC behavior, but a sufficient one. Therefore, while there are alternative explanations of the observed behavior in terms of players' learning \cite{cim14}, we believe that the fact that the outcome of our simulations mixing social and strategic imitation is not far, in behavioral terms, from the experimental results, warrants further research to clarify the explanatory power of this mechanism or similar ones. In this respect, the observed lack of qualitative dependence of the agents' behavior on specific properties of a high dimensional network where the game takes place, being another feature in which both our simulations and experiments coincide, makes such a study even more appealing.

\section{Conclusions}

In this paper, we have provided evidence that social imitation dynamics, added to strategic update rules driving the evolution of players' actions in a PD, leads to consensus. In addition, in an ample range of parameters, the selected consensus is cooperative, even if it is not a Nash equilibrium of the game. In this manner, social imitation given by a VM-like dynamics, helps the individuals of a complex society adopt the most convenient behavior without being influenced by the risks  involved in such a choice, again, at least for not too large punishments. It is important to realize that reaching cooperative outcomes is a product of our mixed dynamics and of the existence of a network, as in a well-mixed population the consensus is always to full defection. The appearance of full cooperation takes place when the punishment $\varepsilon$ for mutual defection is small, as in this case cooperation is not an equilibrium but it is not very far from it, and in that situation social imitation of defectors is not fast enough to induce the other consensus. We also want to stress that this is a  general effect, not limited to the PD, as our earlier study \cite{vrss12} on the totally symmetric coordination game with unconditional imitation shows. Interestingly, in contrast to what takes place in the coordination game, for large $q$ and small $\varepsilon$, the networked PD remains in a disordered dynamic state stable in the  large-size limit and which, in turn, appears to be independent of the topology, at least as far as the heterogeneity of the network is concerned. Moreover, with UI update rule, as in the coordination game we also observe the existence of two regimes in the time evolution towards consensus: power-law when $q$ (the probability of social imitation) is small, and then the main driver of the agents' decision is strategic, and exponential, when $q\rightarrow1$. In this last case, the
system behaves initially according to a pure voter dynamics, reaching a state that is very close to
the final configuration of the voter model. Only after a characteristic time $t^*(q)$ that configuration is in turn affected by
the action of the strategic rule, ending up in a frozen
configuration, with the exception of the region mentioned above.
Naturally, such $t^*(q)$ diverges for $q$ approaching to $1$, where only the voter model regime is left. On the other hand, with REP and MOR only an
exponential decay can be observed (Fig.~\ref{errep-eps0.40}), for every $\varepsilon$ and $q$.

We have also extended our study to other strategic dynamics, in order to assess the mechanisms behind our observations. To that end, we have implemented REP, which copies a neighboring agent's action if it led to better payoff than that of the focal agent, and MOR, which copies a neighboring agent's action with probability proportional to her payoff, and can make mistakes choosing less profitable actions. By comparing these update rules with UI, given by imitation of the best-performing neighbor, we conclude that only if the strategic imitation takes into account the whole neighborhood the payoff-increasing cooperation can arise with some generality. Indeed, with the REP rule consensus to cooperation is the outcome in at most $15\%$ of the realizations, where the asymptotics of MOR is almost random (somewhat biased towards defection) as it is a combination of two bounded rationality imitation rules. Another particular result of UI dynamics is the regime in which the system remains in active state, as with the other two dynamics the final outcome is always full consensus and a frozen configuration. 

It is worth recalling that other researchers have already dealt with models of PD whose dynamics change by tuning a parameter. For example, in~\cite{pin12} the behavior of the game on different topologies is studied as a function of selection pressure, from neutral evolution to pure imitation, while in~\cite{wu13} the robustness of the outcomes of weak selection dynamics in well-mixed populations is analyzed when selection becomes intermediate or strong. There are interesting transitions also in these cases, but we must stress that our study is deeply different: indeed, here the varying parameter is the weight of the VM dynamics, which cannot be considered as a simple noise, since it produces correlations in the system~\cite{vrss12}. Moreover, at variance with reference~\cite{wu13}, it acts on interfaces, that is, it can work only between agents of opposite opinion. 

To conclude in a more general tone, we have found that the mechanism of combined social and strategic imitation is a very powerful one to drive the system towards full consensus, even if the so reached frozen configuration is not an equilibrium of the game as we have seen in this paper. We note that such mechanism does not necessarily lead to desirable states such as cooperation in a networked PD: this is only observed under suitable parameters (in particular, a not too stringent dilemma) and with a ``greedy'' strategic rule such as UI. Otherwise, social imitation leads in the case of the PD to less cooperation than pure strategic behavior. We have also noted that social imitation is not a mutation-type noise, but rather it behaves as interfacial noise as it only acts on active links, where a cooperator and a defector are connected. As such noise, it also leads to enhanced cooperation as compared to pure UI on networks. Finally, we have seen that there a number of features of the experimental results on networked PDs that are recovered from this model, and that a description of those results in terms of transient configurations of our dynamics would be possible. This suggests that our mechanism could provide the basis for alternative explanations of the behavior of human subjects in experiments. 

\section*{Acknowledgments}

D. V. acknowledges support from the PRISMA project (PON04a2 A), within the Italian National Program for Research and Innovation. M. S. M. acknowledges support from grant INTENSE@COSYP (FIS2012-30634) of the Ministerio de Econom\'\i a y Competitividad (MINECO, Spain). J. J. R. receives funding also from the MINECO through the Ram\'on y Cajal program and through the project MODASS (FIS2011-24785). A. S. acknowledges support from grant PRODIEVO from the MINECO. In addition, funding from the EU Commission was received through project LASAGNE.

\end{document}